\documentclass[]{article}
\usepackage{graphicx,longtable,multirow}
\usepackage{times}
\begin{document}

\title{Power Spectral Distribution of the BL Lacertae Object S5 0716+714}

\author{G.R. Mocanu, A. Marcu\thanks{Corresponding author:
alexandru.marcu@phys.ubbcluj.ro}} \date{} \maketitle


\begin{abstract}
Observational data in the BVRI bands of the variable BL Lacertae
Object S5 0716+714 is discussed from the point of view of its
Power Spectral Distribution (PSD). A model of the type $P(f) =
\beta f^{-1}\left [ 1 + \left ( \frac{f} {\delta} \right) ^{\alpha
-1} \right ]^{-1} + \gamma $ is fitted to the data for four null
hypothesis and the Bayesian $p$ parameter for the fits is
calculated. Spectral slopes with values ranging from $1.083$ to
$2.65$ are obtained, with medium values for each band of
$\overline{\alpha} _B =2.028$, $\overline{\alpha} _V = 1.809$,
$\overline{\alpha} _R = 1.932$ and $\overline{\alpha} _I = 1.54$
respectively. These values confirm conclusions of previous
studies, namely that the source is turbulent. Two disk models, the
standard prescription of the Shakura-Sunyaev disk and magnetized
disks exhibiting MagnetoRotational Instability, were discussed. We
found that it is unlikely that they explain this set of
observational data.\newline \textbf{keywords}{turbulence;
                magnetic fields;
                accretion, accretion disks}\end{abstract}

\section{Introduction}
Extensive observational and theoretical efforts have been made in
order to explain IntraDay Variability (IDV) in some classes of
Active Galactic Nuclei (AGN). While variations in
luminosity on scales of an year or more may be explained through
processes usually associated with gravitationally
su\-ppor\-ted Keplerian disks, the significant variations that
occur on a timescale of less than a day are yet unexplained. There
is almost an unanimous consent that explaining variations on all
timescales is equivalent to proposing a robust angular momentum
transport mechanism.

The object BL Lac S5 0716+714 was observed in numerous campaigns
and in different wavelengths and is one of the most manifestly
variable source in the AGN class (Wagner \& Witzel~1995; Wagner et
al.~1996; Qian, Tao \& Fan~2002; Raiteri et al.~2003; Villata et
al.~2008; Poon, Fan \& Fu~2009; Chandra et al.~2011; Carini,
Walters \& Hopper~2011). Flares have been seen in all wavelengths
(Wagner \& Witzel~1995; Poon et al.~2009) and close IDV
correlations between radio (at 6 cm wavelength) and optical (at
650 nm wavelength) have been reported (Wagner et al.~1996; Wagner
et al.~1990; Quirrenbach et al.~1991). Krichbaum et al.~(2002)
discuss 15 years of observations for 40 sources and
report the first detection of mm band IDV for S5 0716+714. We
emphasize on the flaring character, as no or little
evidence for periodicity has been found. Qian~(1995) reported that
behavior changed from quasi periodic daily to less periodic weakly
oscillations and Quirrenbach et al.~(1991) report transitions from
one dominant IDV scale to another. Krichbaum et al.~(2002) find
that above 8GHz the variability index increases with frequency.
Qian et al.~(2006) report this object in a study of IDV sources
with very high polarizations. This indicates the presence of
uniform background magnetic fields in the source.

When such a wealth of observational data is at hand, one may use
it to discriminate between theoretical models (as e.g. Kraus et
al.~1999). Quirrenbach et. al.~(1992) comment that the
correlated variations in simultaneous optical and radio
variability cannot be explained by the action/interaction with the
InterStellar Medium (ISM). Qian et al.~(1996a;1996b) comment that
shock propagating in an oscillatory jet might explain IDV and
correlation between radio and optical IDV. Kirk \&
Mastichiadis~(1992) propose models based on injection and
acceleration of particles. Be\-gel\-man, Rees \& Sikora~(1994)
refines the relativistic jet model to explain the high brightness
temperature theoretically associated to IDV. The model
successfully reproduces the observed spectral index
variations.

Chandra et al.~(2011) discuss variability in optical BVRI bands during a 5 day
monitoring campaign in March 2010. They present light curves and
calculate variation rates. The fast variations and the high
amplitudes in magnitude are difficult to explain through accretion
disk models. If variation in the Doppler factor is allowed, the
shock in jet framework might explain the bluer when brighter
behavior, but it cannot explain the microvariability (i.e.
variability on timescales of a few tens of minutes).

Carini et al.~(2011) report B and I bands microvariability for a
5 night observation campaign in March 2003. They perform light curve analysis,
timescale analysis, color analysis, structure function analysis
and cross correlation analysis. They firmly reject the hypothesis
that the observed spectra might arise following electron cooling.
Their conclusion is that the observed microvariability is the
result of a fractional noise process, i.e. the source of the
variations is a turbulent process.

Azarnia, Webb \& Pollock~(2005) analyze a set of 10 nonconsecutive
R band light curves by using the Discrete Fourier Transform in
order to obtain the possible noise characteristics of the time
series. The results they obtained led them to speculate that
microvariability is the result of complex turbulent relativistic
plasma process.

A very interesting type of IDV analysis is based on the
calculation of the fractal dimension of the light curves (Leung et
al.~2011b). The fractal dimension of the R-band observations
indicates an almost pure "Brownian noise" (random walk)
spectrum.

The purpose of this paper is to discuss the observational data
first presented in Poon et al.~(2009) from the point of view of
its PSD. After presenting (Section~\ref{sect:observations}) the
observational data, a
detailed PSD analysis of the variability is performed
(Section~\ref{sect:discussion}). An attempt is made to fit two
accretion disk models to the data (Section~\ref{sect:models}).

\section{Observational data\label{sect:observations}}
The data we consider has been recorded in the optical band (more
precisely, the BVRI bands) during October and December 2008 and
February 2009. These sets of data and the observational technical
characteristics have been thoroughly analyzed and dis\-cussed in
Poon et al.~(2009). There are compelling arguments that the source is
variable in the BVRI band and that the flares at different
wavelengths are due to the same generating mechanism.

The analysis in Poon et al.~(2009) includes the spectral changes this
source exhibits, i.e. the way in which the amplitude changes as a
function of wavelength, which is equivalent to the Spectral Energy
Distribution (SED). We wish to continue their work by introducing
Power Spectral Distribution (PSD) analysis in all available
wavelengths.

The observational data is presented in
Table~\ref{table:variabilityData}, where the columns have the
following meaning

\begin{enumerate}

\item{}identification code for each band and each Julian Day of
observations (i.e. R5 means "about data taken in the R band in the
fifth date");

\item{}the actual Julian Date. It may be that some observations
were made from 2454865.99 to 2454866.4 so we considered them as
being part of the same day and included them in the analysis as
such;

\item{}band, from B (blue, $\lambda _B$ $= 440nm$), R (red, $\lambda
_R$ \newline $= 630 $$nm$), V (visible, $\lambda _V = 550 nm$), I (infrared,
$\lambda _I = 900 nm$);

\item{}amplitude of variability for each day and for that specific
band (Poon et al.~2009), calculated here in units of $\sigma$

\begin{equation}
A = \frac{\sqrt{\left(A_{max}-A_{min}\right)^2-2\sigma
^2}}{\sigma},\label{eq:defA}
\end{equation}
where $\sigma$ will be given below;

\item{}number $N$ of data points in that Julian Day (JD) and for
that specific band;

\item{}$\overline{m}$, the medium magnitude measured that day\newline
$\overline{m} = \frac{\sum _{i=1}^N m_i}{N}$, where $m_i$ stands
for the magnitude at one point, $i=\overline{1,N}$;

\item{}root mean square deviation error $\sigma$ calculated as $N
\sigma = \sqrt{\sum _{i=1}^N \left(m_i^2 - \overline{m}^2\right)}$
for each day and in that specific band.

\end{enumerate}

\section{PSD Analysis\label{sect:discussion}}

From the light curves (Poon et al.~2009) and the values
of the variability amplitudes (the $A$ values in
Table~\ref{table:variabilityData}) it is obvious that this BL Lac
object presents microvariability in the BVRI bands. Our purpose is
to determine the slope of the power spectrum of the variations. To
this end, the \emph{software R} and the \emph{bayes.R} script are
used, designed to detect periodic signals in red noise
(Vaughan~2010). Periodicity is not expected, but the software is
useful in obtaining fits of the slope of the power spectrum.

We will do this for each day of observations. The theoretical
working models used by the .R routine (i.e. the available null
hypotheses) are power law plus constant:

\begin{equation}
H_0: S(f) = \beta f^{-\alpha} + \gamma,\label{eq:H0}
\end{equation}
bending power law plus constant:

\begin{equation}
H_1: S(f) = \beta f^{-1} \left [ 1 + \left ( \frac{f}{\delta}
\right ) ^{\alpha - 1} \right ] ^{-1} + \gamma,
\end{equation}
power law:

\begin{equation}
H_2: S(f) = \beta f^{-\alpha},\label{eq:H2}
\end{equation}
and bending power law:

\begin{equation}
H_3: S(f) = \beta f^{-1} \left [ 1 + \left ( \frac{f}{\delta}
\right ) ^{\alpha - 1} \right ] ^{-1}.\label{eq:H3}
\end{equation}

After running the script for all the days and in each band,
Table~\ref{table:spectral} was obtained, where the columns have the
following significance

\begin{enumerate}

\item{}observation day, as defined in
Table~\ref{table:variabilityData};

\item{}model used, i.e. one of the four available null hypotheses
$H_i$ (Eq.~\ref{eq:H0}-\ref{eq:H3});

\item{}values of parameter $\theta _1 = \alpha$ (the standard
deviation is given between the square brackets);

\item{}values of parameter $\theta _2 = \ln{\beta}$ (the standard
deviation is given between the square brackets);

\item{}values of parameter $\theta _3 = \ln{\gamma}$ (the standard
deviation is given between the square brackets). For the $H_2$
hypothesis there is no third parameter and this was denoted by a
$-$ symbol in the appropriate place. When there is no entry for a
$H_i$ it means that for that specific case the software returned
an error;

\item{}values of parameter $\theta _4 = \ln{\delta} $ (the
standard deviation is given between the square brackets);

\item{}the (Bayesian) posterior predictive p-value is used for
model checking and has the advantage of having no dependence on
unknown parameters. It may be used to assess whether the data are
consistent with being drawn from the model (Vaughan~2010). If the
values of the statistics are very small it is unlikely that the
proposed model could reproduce the data.

\end{enumerate}

\begin{figure}
   \centering
   \includegraphics[width=8cm]{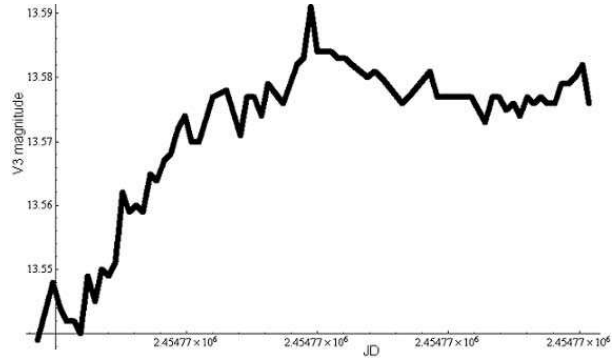}
      \caption{V3 time series.}
         \label{fig:V3TimeSeries}
   \end{figure}

\begin{figure*}
   \centering
   \includegraphics[width=6cm]{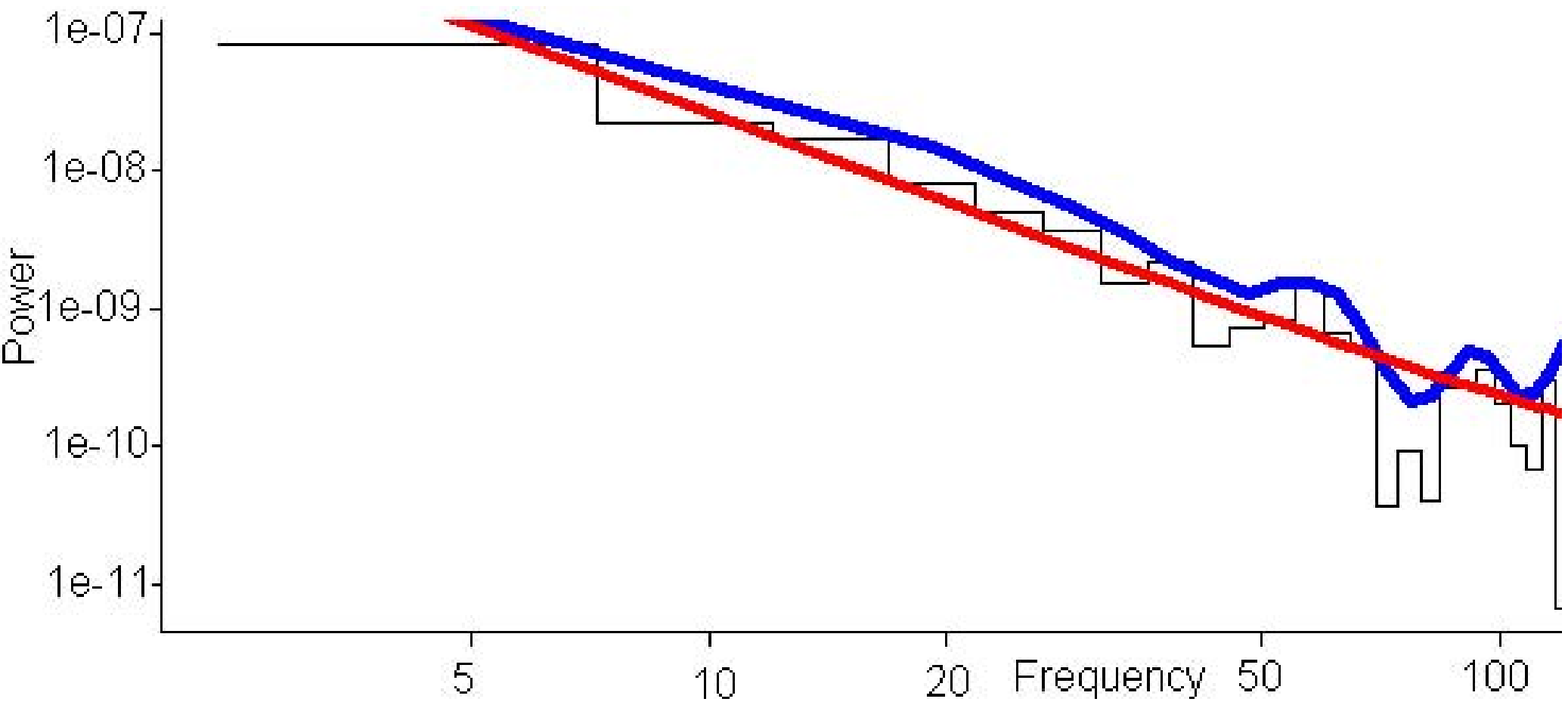}
   \includegraphics[width=6cm]{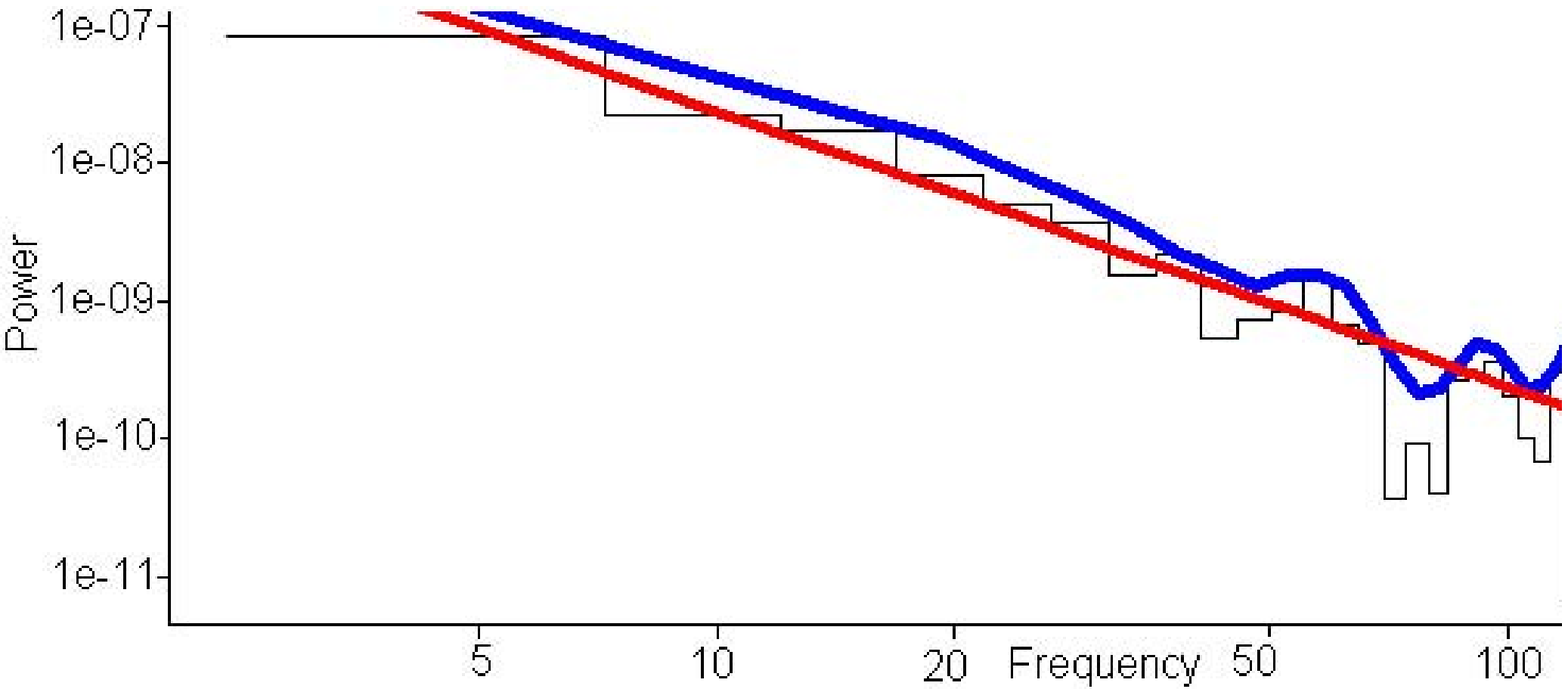}
   \caption{Fit (red curve) for model $H_0$ (left) and model $H_2$
(right) applied to the PSD (black curve) of the V3 time series.}
          \label{fig:V3H0PohF}
    \end{figure*}

\begin{figure*}
   \centering
   \includegraphics[width=6cm]{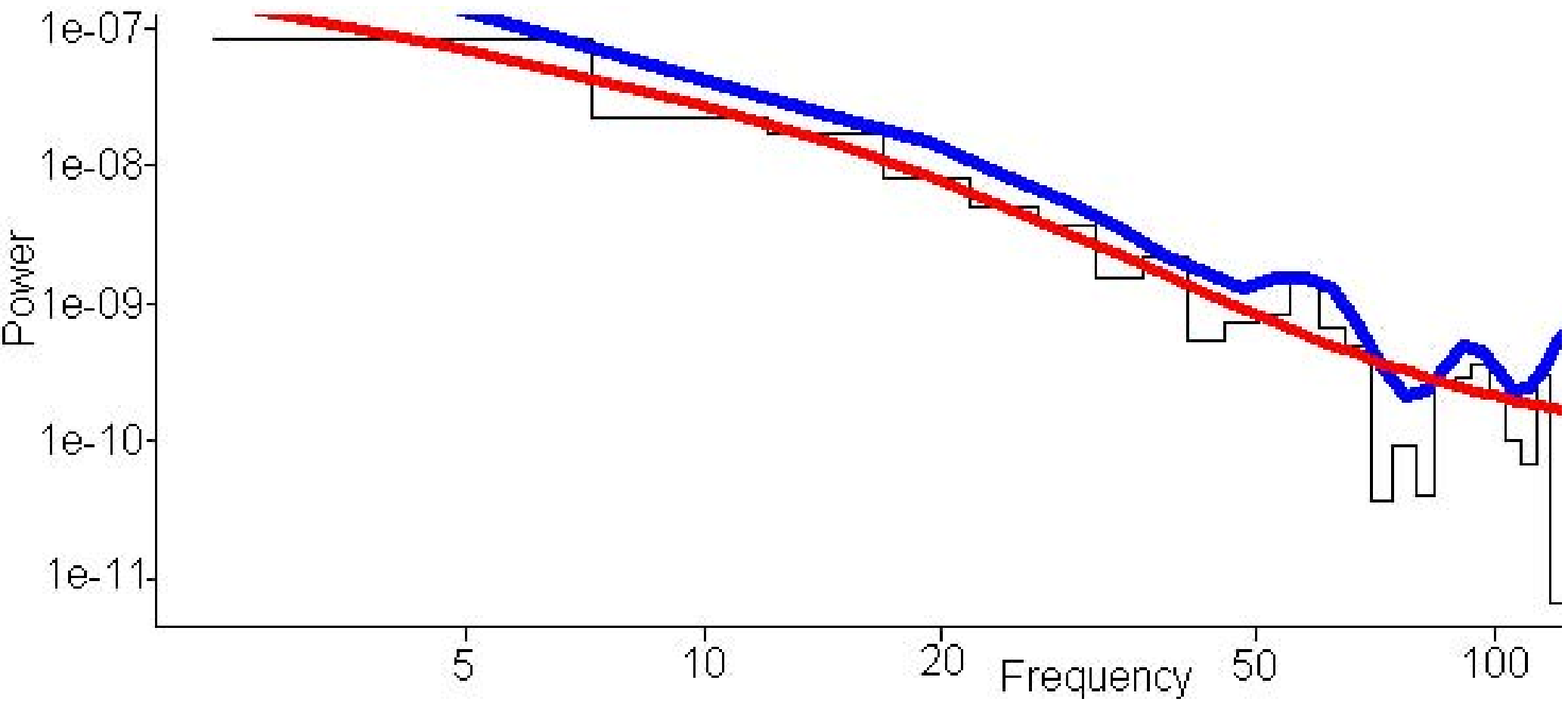}
   \includegraphics[width=6cm]{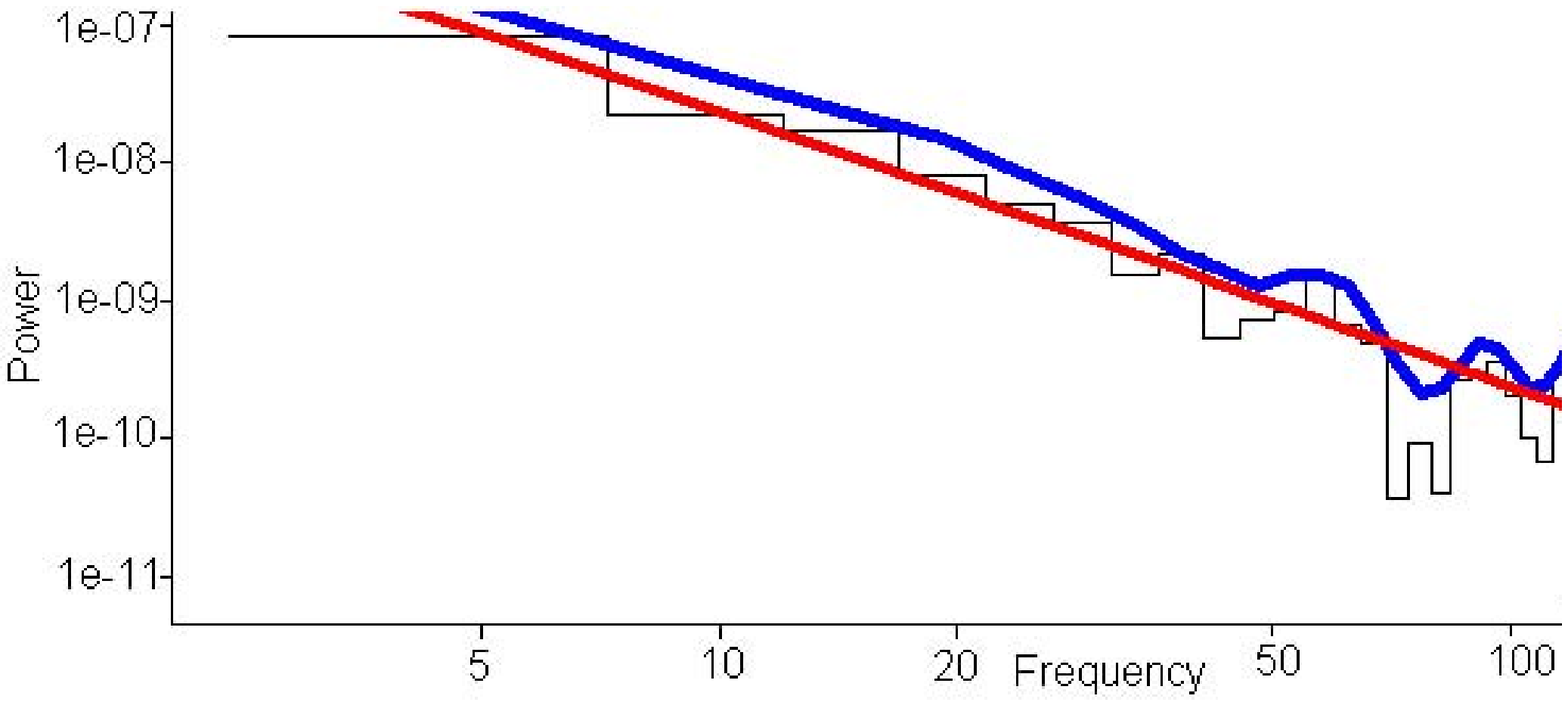}
   \caption{Fit (red curve) for model $H_1$ (left) and model $H_3$ (right) applied to the PSD (black curve) of the V3 time series.}
          \label{fig:V3H1PohF}
    \end{figure*}

As an example of the fits, the time series for V3
(Fig.~\ref{fig:V3TimeSeries}) and the power spectra fit for models
$H_0$ (Fig.~\ref{fig:V3H0PohF} left) and $H_2$
(Fig.~\ref{fig:V3H0PohF} right) and models $H_1$
(Fig.~\ref{fig:V3H1PohF} left) and $H_3$ (Fig.~\ref{fig:V3H1PohF}
right) are shown here.

Some comments are in order regarding the results of the
PSD analysis. First, the bending power law null hypotheses ($H_1$
and $H_3$) fail to produce results in most of the cases. As seen
in Table~\ref{table:spectral}, for the cases where they do produce
results, the standard mean deviations of the parameters are a lot
bigger than for $H_0$ and $H_2$.

Secondly, while the correct interpretation of the parameter set
$\{\theta _2, \theta _3, \theta_4\}$ can provide valuable
information, we will be interested in the PSD slope, i.e. $\theta
_1 \equiv \alpha$.

Our interest follows from the known fact
that the numerical values of the slopes of the PSD provide insight
to the nature of the mechanism leading to the observed
variability. In statistical analysis, if $X$ is some fluctuating
quantity, with mean $\mu$ and variance $\sigma ^2$, then a
correlation function for quantity $X$ is defined as

\begin{equation}
R(\tau) = \frac{\langle \left ( X_s - \mu \right ) \left (
X_{s+\tau} - \mu \right) \rangle}{\sigma ^2}.
\end{equation}
where $X_s$ is the values of $X$ measured at time $s$ and $\langle
\rangle$ denotes averaging over all values $s$. The Power Spectral
Distribution is defined based on the correlation function as

\begin{equation}
P(f) = \int _{-\infty} ^{+\infty} R(\tau) e^{-\imath 2\pi f \tau}
d\tau
\end{equation}
and it is straightforward to see its importance in terms of the
"memory" of a given process. For example, if $X$ is the B band
magnitude of the disk, the slope of the PSD of a time series of
$X$ provides insight to the degree of correlation the underlying
physical process has with itself. The system needs additional
energy to fluctuate and this mechanism is historically best
explained for Brownian motion, in which case the energy is
thermal. Brownian motion produces a PSD $P(f) \sim f^{-2}$.
Completely uncorrelated evolution of a system produces white
noise, with a PSD $P(f) = f^0 = const.$ It is then very
interesting to try and explain how does a system evolve so as to
produce a PSD for which $\alpha$ is neither $0$ nor $2$, as is the
case for the time series discussed in this paper.

With this in mind, we now look
at the $\theta _1$ and $p_B$ co\-lumns from
Table~\ref{table:spectral}. From each observation day we want to
emphasize on the value of the PSD slope which satisfies both
minimum standard mean deviation and maximum $p_B$ criteria.
However, this is not the case for all entries in the Table. In
order to choose a value for the spectral slope (written in
boldface in the Table) the following guidelines were used

\begin{enumerate}

\item{}we choose the values which clearly satisfy both the
criteria and $p_B>0.5$ (12 time series);

\item{}for cases when all $p_B$ lie in the interval $[0.8,1]$, but
the minimum standard deviation is exhibited by the model with
lower $p_B$, we favor the minimum standard mean deviation
criterion (11 time series);

\item{}for cases when one $p_B$ is above $0.5$ and the rest are
below, we favor the maximum $p_B$ criterion (4 time series);

\item{}for cases when all $p_B$ are below 0.5 we consider that the
source does not behave like the null and do not use the obtained
spectral slope for any further calculation. From our 41
time-series, 8 lie in this category (identification code written
in bold);

\item{}for time series which fall in neither category we favor the
maximum $p_B$ criterion (6 time series). This would be the case
of, e.g., V9.

\end{enumerate}

With these
considerations, the data provides slopes with values ranging from
$1.083$ to $2.65$, with medium values for each band of
$\overline{\alpha} _B =2.028$, $\overline{\alpha} _V = 1.809$,
$\overline{\alpha} _R = 1.932$ and $\overline{\alpha} _I = 1.54$
respectively.

\begin{figure*}
   \centering
   \includegraphics[width=6cm]{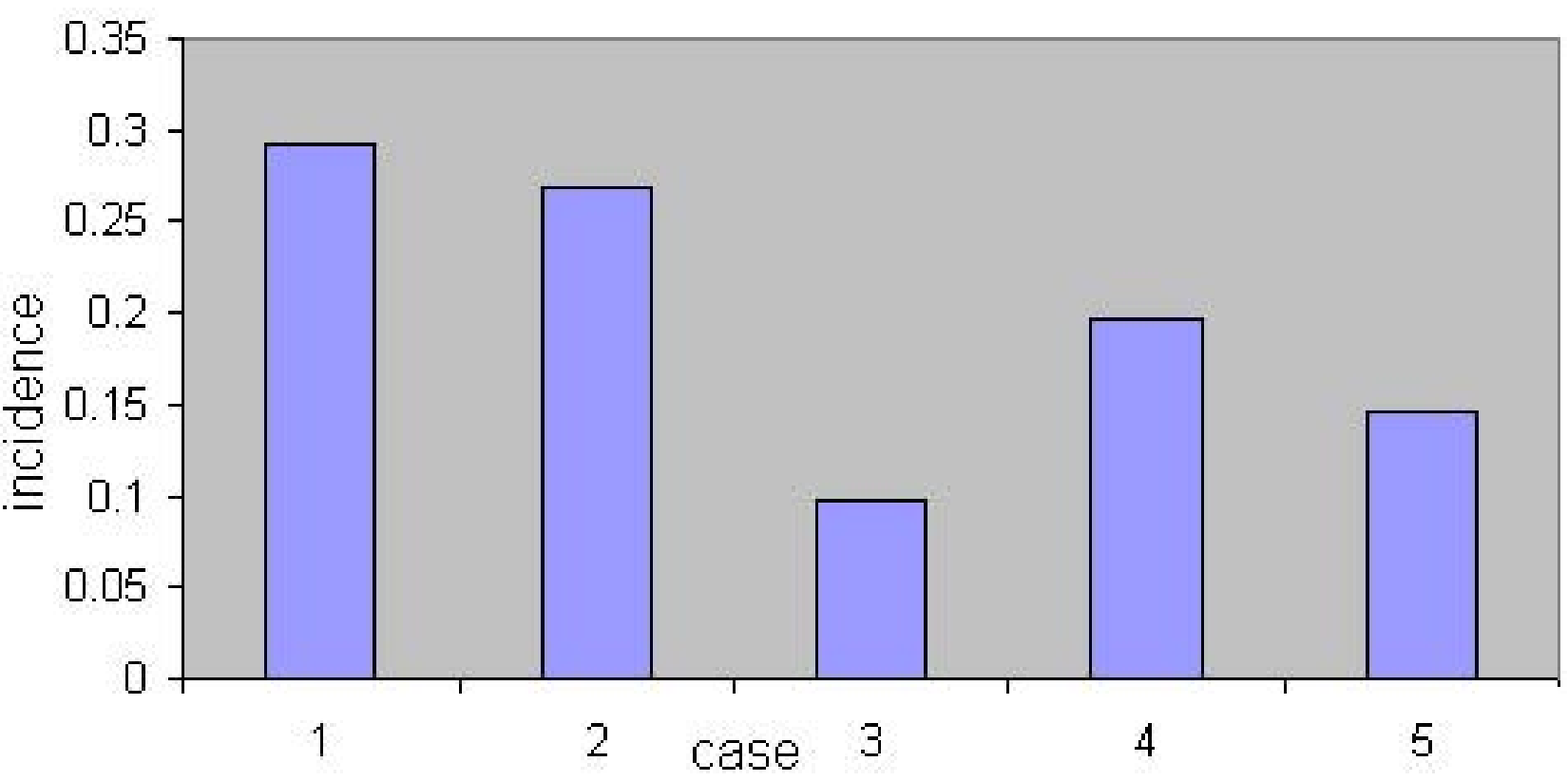}
   \includegraphics[width=6cm]{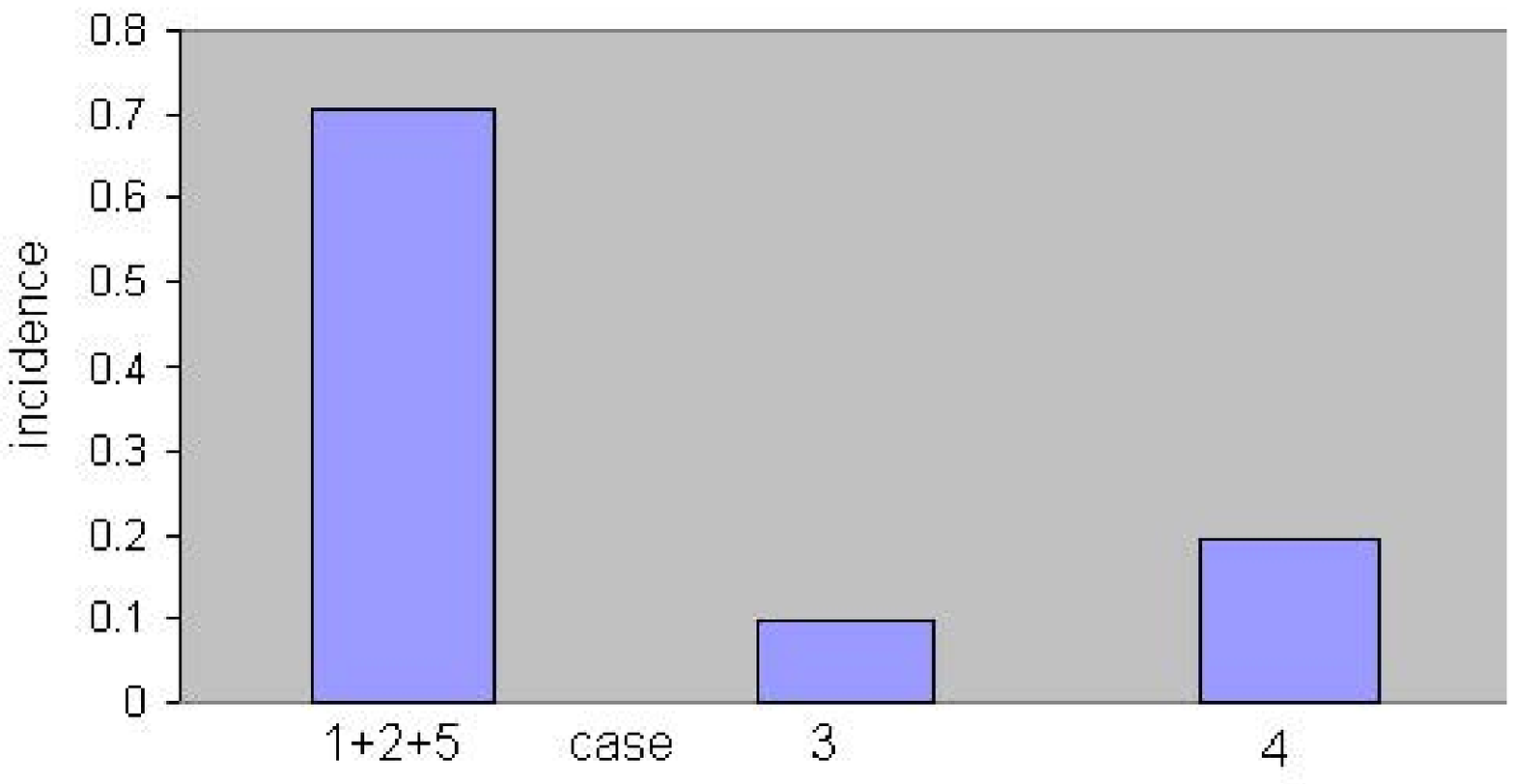}
    \caption{Left: Histogram of the number of time-series acceptably fitted by the null
hypotheses of the .R software for the five cases discussed in the
text. Right: Histogram for the occurrence of a valid power-law fit
when cases 1,2 and 5 are taken together.}
          \label{fig:hist1}
    \end{figure*}

It might prove to be an interesting exercise to do a histogram of
these values (Fig.~\ref{fig:hist1} left) to have a visual
description of the validity of the power-law behaviour of the
PSD.
This is a pretty good result, showing that at least for this data
set we may consider that the PSD behaviour of the source is well
fitted by a power-law. This is a conclusion that becomes even more
clearer if one views cases 1,2 and 5 as one group and updates the
histogram as in Fig.~\ref{fig:hist1} right.

For the same object and a set of 10 nonconsecutive time series for
the R band (2003 to 2005), Azarnia et al.~(2005) obtain values for
the spectral slope between $-0.9 \pm 0.122$ and $-1.393 \pm
0.1005$. As emphasized by the authors, the $1/f$ results are not
conclusive, but it is clear that the process noise is not white
noise.

\section{Theoretical models\label{sect:models}}

Historically, there have
been attempts to explain the variability through external effects
e.g. RISS (Refractive Scintillation in the interstellar medium)
(Wambsganss et al.\newline1989), microlensing (Wagner \&
Witzel~1995) or based on source morphology, e.g. a cluster of
independently radiating objects (Krolik~1999, page 76). However,
most of them fail because they do not predict the entire range of
effects associated with variability.

A number of models study variability in the framework of efficient
angular momentum transfer within the accretion disk, assuming that
perturbations in this mechanism are responsible for the IDV.
Mechanisms of angular momentum transfer may be conceptually
divided in three different classes (Papaloizou \& Lin~1995) based
on the fundamental behaviour of the disturbance: hydromagnetic
winds, waves in disks mechanism and thermal convection. None
completely reproduces the observed characteristics of IDV.

Through a series of papers (Mineshige, Ouchi \& Nishimori~1994a;
Mineshige, Takeuchi \& Nishimori~1994b; Yo\-ne\-ha\-ra, Mineshige
\& Welsh~1997) there was an attempt to reproduce the PSD
characteristics of IDV in a Self Organized Criticality framework.
Realistic PSDs for the high energy (X-Ray) part of the spectrum
may be obtained in this way.
\newline \newline
\textbf{Magneto Rotational Instability}
\newline \newline
The MagnetoRotational Instability (MRI, Balbus \& Hawley~1991) is
the most promising mechanism yet, its strength residing in the
combination of differential rotation and the presence of an
initially weak magnetic field. This approach has been
systematically developed in the last few years to include
theoretical and numerical discussion of various mag\-netic field
configurations both in the linear and nonlinear regimes (Balbus \&
Hawley~1991,~1992a,~1992b; Hawley \& Balbus~1991,~1992; Hawley,
Gammie \& Balbus~1995).

   \begin{figure*}
   \centering
   \includegraphics[width=6cm]{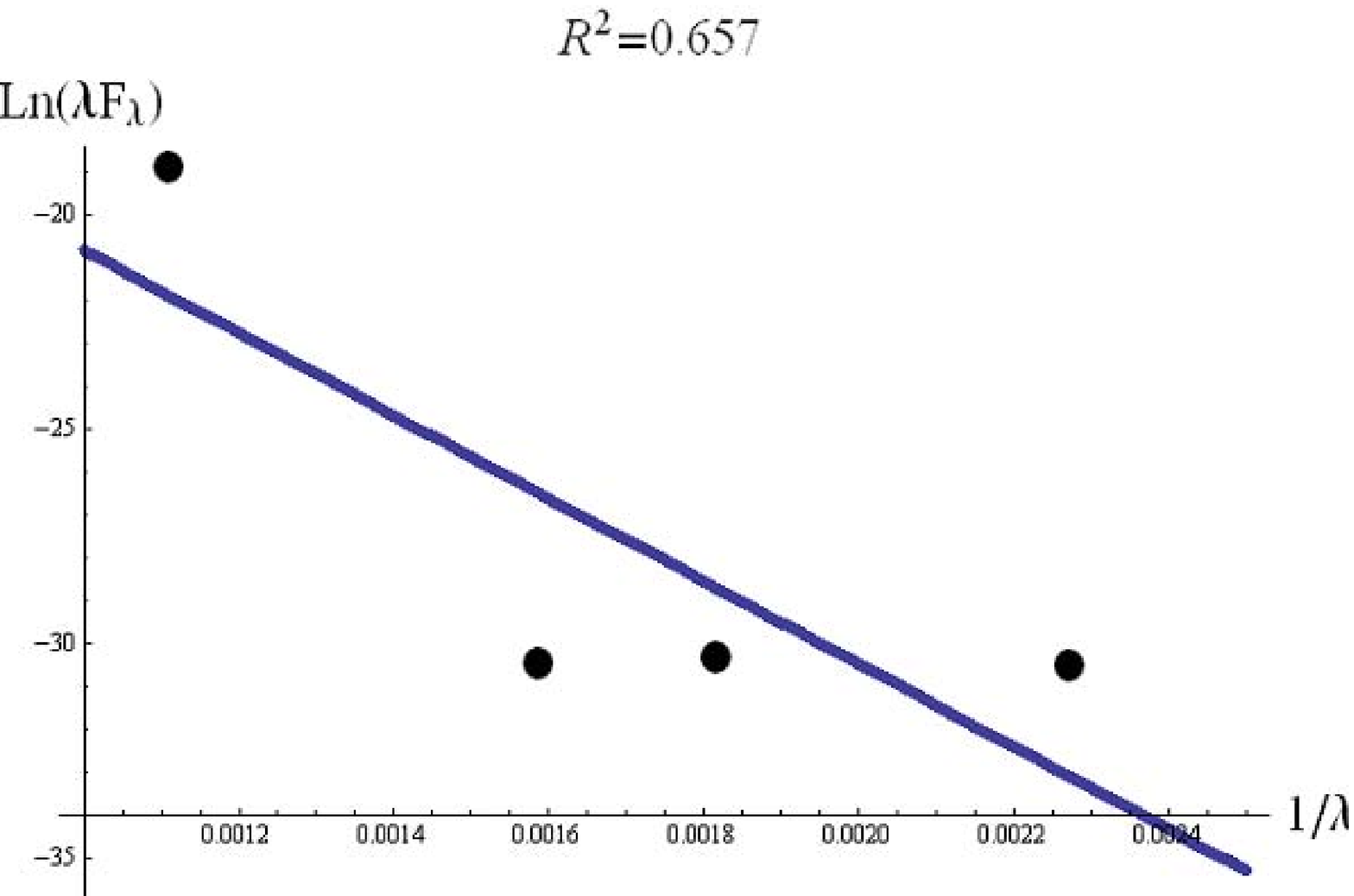}
   \includegraphics[width=6cm]{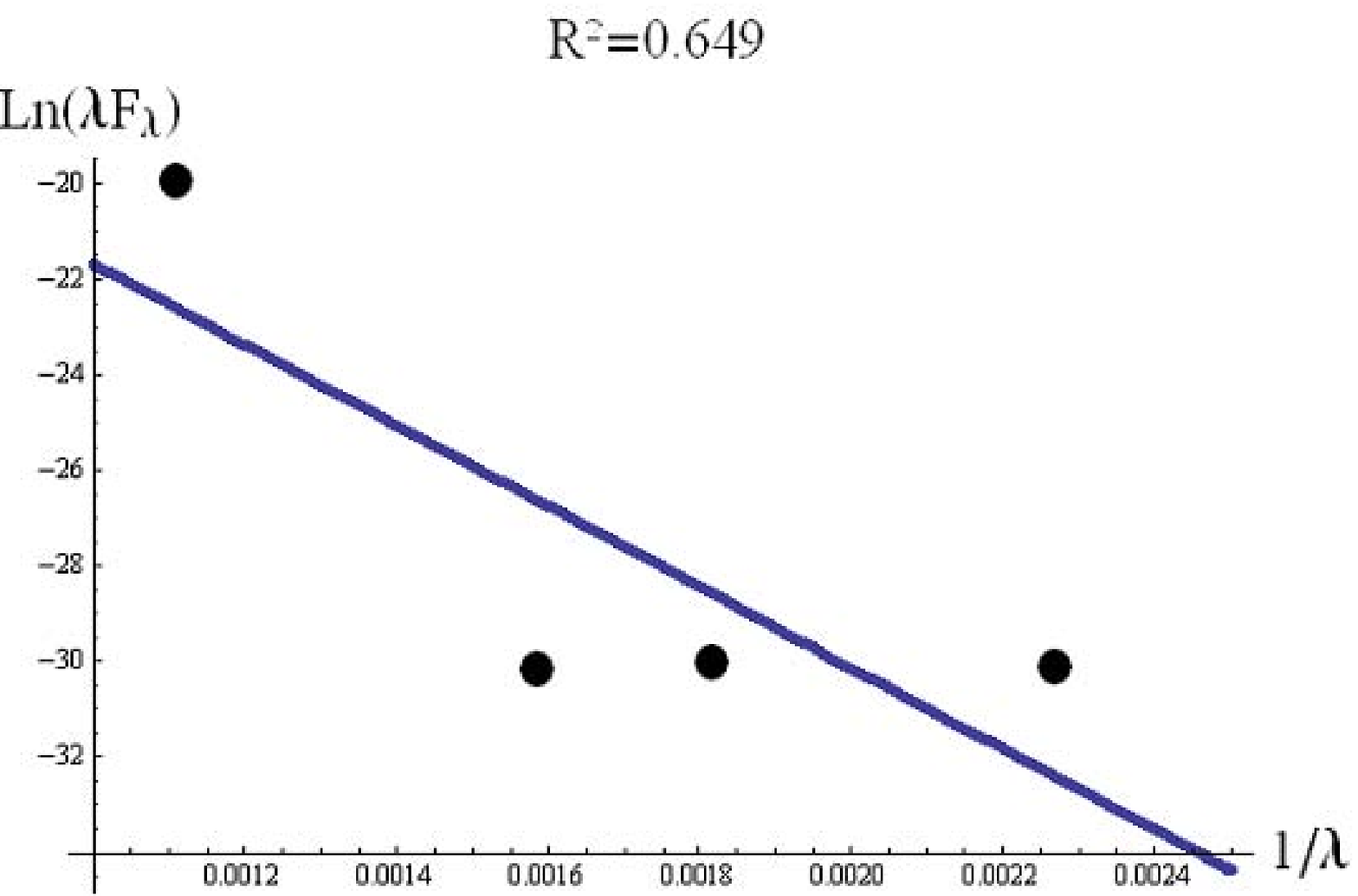}
   \caption{Plot of the logarithm of the medium $\lambda F_\lambda$ as a
function of $\lambda ^{-1}$, for the entire observational period
(left) and for JD 2454866 (right). A fit of the type $\ln{\lambda F_\lambda} =
a/\lambda + b$ was attempted, but it provides unsatisfactory
results.}
          \label{fig:magMed}
    \end{figure*}

In order to qualitatively test the effect of MRI onset on the
emergent spectrum, we propose the following reasoning.
Theoretically, the emergent spectrum is proportional to the first
power of the Reynolds-Maxwell stress tensor (Blaes~2002). The
definition of this stress tensor, in the context adopted by us
here is given in Balbus \& Hawley~(1998). Its value presents a
dependency of the type $\exp\{-3/\lambda\}$. It is further assumed
that $\lambda$, the wavelength of the disturbance, is also the
observed wavelength. We calculated a medium flux for each
wavelength for the entire observational campaign, and plotted $Ln\left(\lambda F_\lambda\right)$ as a function of the frequency corresponding to
observed wavelength (Fig.\ref{fig:magMed} left).
The same algorithm was followed for a day of observations where
data were available in all filters, i.e. for JD 2454866
(Fig.~\ref{fig:magMed} right). In both cases, according to theory,
we would expect a linear dependency, of the type $y \sim -3x$
which was not found. In fact, following this simple analysis, not
even the linear character of the dependency is confirmed. An
attempt to fit a function $y = ax+b$ to the data produced $R^2 =
0.657$ for the entire observational period and $R^2 = 0.649$ for
JD 2454866.
\newline \newline
\textbf{The Shakura Sunyaev disk}
\newline \newline
The disk model presented by Shakura and Sunyaev
in a series of papers (Shakura \& Sunyaev~1973;~1976) starts from
fundamental equations for geometrically thin disk accretion and
perturbs these equations. They solve for the perturbations in
surface density and height and express the total luminosity of the
disk in terms of these perturbations. If the scale of perturbation
is quantified by $\Omega$ ($\tau \propto 1/\Omega$), $\Omega$
evolves through a small strip of the parameter spa\-ce (Sha-ku\-ra
\& Sunyaev~(1976), Fig. 1), $\Omega / (\alpha _{SS} \omega) \in
[0.02,0.2]$, where $\alpha _{SS}$ is the Sha\-ku\-ra-Su\-ny\-aev
coefficient and $\omega = \sqrt{GM/R^3}$ is the Keplerian angular
frequency.

If adimensional parameters $b=M/M_\odot$ and $d = R/R_g$ are used,
the constraint from Shakura \& Sunyaev~(1976) may be re-written as

\begin{equation}
\frac{d^{3/2}b}{\tau \alpha _{SS}} \in [0.404, 4.047]\cdot
10^4.\label{eq:const}
\end{equation}

For generally accepted numerical values for super-ma\-ssive black
holes, i.e. $d = 10$ and $b=10^9$, and considering that the
variability timescale is four hours, the constraint becomes

\begin{equation}
\alpha \in [0.18, 1.72] \cdot 10^{-2}.\label{eq:accepted}
\end{equation}

Numerical simulations for the coefficient $\alpha$ place its
va\-lue somewhere around $10^{-2}$. The result in
Eq.\ref{eq:accepted} can be considered as a success of the model.
It was already known that this mathematical formalism works but
the problem still remains why it works, i.e. put the value of the
coefficient of firm physical grounds.

However, based on recent data (Fan et al.~2011)\footnote{We thank
J. Fan for pointing this out.} for this object,  $\tau = 216 s$
and $b \in [10^{7.68},$ $ 10^{8.38}]$. If $d=5$, then $\alpha
_{SS} = 1.82$ for $b=10^{7.38}$ and the value of $\alpha _{SS}$
grows as $b$ and $d$ grow (Eq.~\ref{eq:const}). It is then quite
clear that the $\alpha _{SS}$ prescription  of the standard disk
model cannot explain the variability reported by Fan et
al.~(2011). With the benefit of this hindsight, we make the
following argumentation, in order to obtain a "rule of thumb" to
quickly asses whether or not a set of observational data may be
explained by the standard disk model. Starting from
Eq.~\ref{eq:const} and considering $d\in [5,10]$ a relation
between $\tau$, $b$ and $\alpha _{SS}$ is obtained

\begin{equation}
\alpha _{SS} \in [0.0825, 2.475]\cdot \frac{b}{\tau} \cdot
10^{-4}.
\end{equation}

If validity of the standard model is assumed, we must impose
$\alpha _{SS}<1$, which means that the ratio of the black hole
mass to the variability time scale should be saturated at a finite
value

\begin{equation}
\frac{b}{\tau} < [0.04,1.212] \cdot 10^5.\label{eq:thumb}
\end{equation}

This rule is very restrictive if one considers that typical values
for $b$ for AGNs are of the order $10^7 - 10^{10}$ and that IDV
refers to timescales too small compared to those needed to satisfy
relation~\ref{eq:thumb}.
\newline \newline
\textbf{Modelling IDV in a stochastic turbulence framework}
\newline \newline
A stochastic process is defined as a process where one or
more of the variables of interest (called random variables) have
some degree of uncertainty in their realization. At its base,
stochastic modelling analyzes the evolution of some random
variable and of its distribution function, with the aid of
Langevin type equations and the Focker-Planck equation. Efforts to
explain IDV in this framework are being developed. Leung et
al.~(2011a) take the random variable as the height of the disk and
the stochastic component of the Langevin equation is set to mimic
the interaction of the disk with a background cosmic environment.

If the random variable is taken to be the magnitude in one
of the BVRI bands, three types of analysis of observational data,
namely structure function analysis (Carini et al.~2011), fractal
dimension analysis (Leung et al.~2011b) and DFT analysis (Azarnia
et al.~2005) plus our own analysis establish that the
source is turbulent, i.e. there is "intrinsic" noise superimposed
on the deterministic behaviour of the source. The actual nature,
onset and dissipation of turbulence is still a topic of discussion
(Shakura \& Sunyaev~1976; Balbus \& Hawley~1998). At this point
one can only speculate, but one educated guess is that the
stochastic reconnection process proposed by Lazarian \&
Vishniac~(1999) is an important part in producing of IDV. This
process has proven successful in explaining fast and energetic in
events over a large range of lengthscales.

\section{Conclusions\label{sect:conclusions}}

The spectral slope
and Bayesian confidence p-parameter for the BVRI bands
observational data (Poon et al.~2009) were calculated. This was
done for four null hypothesis a\-va\-i\-la\-ble in the .R software
(Vaughan~2010), Eqs.~\ref{eq:H0}-\ref{eq:H3}. The mean values for
the spectral slope are $\overline{\alpha} _B =2.028$,
$\overline{\alpha} _V = 1.809$, $\overline{\alpha} _R = 1.932$ and
$\overline{\alpha} _I = 1.54$. A histogram of the number of
time-series which are well fitted by a PSD power-law is very
encouraging, showing that the source presents power-law behavior
in the BVRI bands. The values of the spectral coefficient confirm
previous results which state that the source is noisy in a
nontrivial way (Leung et al~2011b; Carini et al.~2011; Azarnia et
al.~2005).

An attempt was made to explain the data in the context of two
accretion disk models, the Shakura-Sunyaev disk and a magnetized
disk exhibiting MRI. For standard AGN parameters the effective
Shakura-Sunyaev parameter, $\alpha _{SS}$, is within the
theoretically correct interval, i.e. smaller than 1. However, if
the new observational data of Fan et al.~(2011) is taken into
account, the hypothesis that IDV is produced within the disk is
clearly not valid, since it would produce an $\alpha _{SS}$ with
values well above $1$. A naive rule of thumb to quickly assess
wether or not IDV exhibited by some timeseries is produced within
the standard prescription is derived, Eq.~\ref{eq:thumb}. The
attempt to fit the data within an MRI framework was also
unsuccessful, Fig.~\ref{fig:magMed}.

\subsubsection*{Acknowledgements} G.M. would like to thank T. Harko
for valuable discussions on the topic of variability and
suggestions regarding this work. This work was possible with the
financial support of the Sectoral Operational Programme for Human
Resources Development 2007-2013, co-financed by the European
Social Fund, under the project number POSDRU/107/1.5/S/76841 with
the title "Modern Doctoral Studies: Internationalization and
Interdisciplinarity". The authors thank the anonymous referee for
valuable comments and suggestions.

\newpage


\begin{onecolumn}
\begin{longtable}[!t]{|p{0.6cm}| p{1.3cm}|p{0.3cm}|p{1.2cm}|p{0.8cm}|p{1.5cm}|p{1.5cm}|}\hline
    & JD & & A[$\sigma$] & N & $\overline{m}$[magn] & $\sigma$ \\ \hline
    B1 & 2454824 & B & 43.28191 & 107 & 14.30222 & 0.003395 \\ \hline
    B2 & 2454826 & B & 45.39896 & 127 & 14.25919 & 0.001805 \\ \hline
    B3 & 2454828 & B & 19.18941 &  37 & 14.55843 & 0.003118 \\ \hline
    B4 & 2454866-2454867 & B & 55.58283 &  126 & 13.8741 & 0.003111 \\ \hline
    B5 & 2454871-2454872 & B & 35.77494 &  103 & 14.24217 & 0.007821 \\ \hline
    B6 & 2454872-2454873 & B & 35.76081 &  86 & 14.6071   & 0.00366 \\ \hline

    V1 & 2454765 & V & 27.03456 &  93 & 13.53289 & 0.007609 \\ \hline
    V2 & 2454766 & V & 26.14296 &  82 & 13.50011 & 0.006226 \\ \hline
    V3 & 2454767 & V & 57.34061 &  64 & 13.57052 & 0.000907 \\ \hline
    V4 & 2454770 & V & 30.20165 &  80 & 13.73929 & 0.004498 \\ \hline
    V5 & 2454824 & V & 49.68765 & 107 & 13.80678 & 0.002776 \\ \hline
    V6 & 2454825 & V & 32.8199 & 108 & 13.65370 & 0.003988 \\ \hline
    V7 & 2454826 & V & 51.5246 & 128 & 13.76498 & 0.001474 \\ \hline
    V8 & 2454828 & V & 38.47998 &  52 & 14.06906 & 0.001948 \\ \hline
    V9 & 2454829 & V & 51.44625 & 148 & 13.83567 & 0.00136 \\ \hline
    V10& 2454830 & V & 46.83345 &  85 & 13.78665 & 0.00064 \\ \hline
    V11& 2454865-2454866 & V & 53.95782 & 178 & 13.77554 & 0.002001 \\ \hline
    V12& 2454866-2454867 & V & 51.84826 & 127 & 13.4035 & 0.002873  \\ \hline
    V13& 2454871-2454872 & V & 34.44135 & 103 & 13.75483 & 0.007485 \\ \hline
    V14& 2454872-2454873 & V & 35.6525 &  86 & 14.1083 & 0.003335 \\ \hline

    R1 & 2454765 & R & 26.80611 &  92 & 13.11834 & 0.006743 \\ \hline
    R2 & 2454766 & R & 27.18165 &  84 & 13.09189 & 0.005658 \\ \hline
    R3 & 2454767 & R & 25.38848 &  62 & 13.15161 & 0.001652 \\ \hline
    R4 & 2454770 & R & 27.59172 &  79 & 13.31928 & 0.00409  \\ \hline
    R5 & 2454824 & R & 49.72761 & 107 & 13.36121 & 0.002493 \\ \hline
    R6 & 2454825 & R & 37.45431 & 111 & 13.21392 & 0.003629 \\ \hline
    R7 & 2454826 & R & 46.4458  & 127 & 13.34012 & 0.001485 \\ \hline
    R8 & 2454828 & R &   22.699 &  51 & 13.61788 & 0.002462 \\ \hline
    R9 & 2454829 & R & 41.44746 & 148 & 13.40384 & 0.001591 \\ \hline
    R10& 2454830 & R & 47.92965 &  85 & 13.35033 & 0.000667 \\ \hline
    R11& 2454865-2454866 & R & 55.61234 &  177 & 13.35199 & 0.002013 \\ \hline
    R12& 2454866-2454867 & R & 47.78895 & 110 & 12.99796   & 0.002845 \\ \hline
    R13& 2454871-2454872 & R & 32.6797 & 102 & 13.33697 & 0.007184 \\ \hline
    R14& 2454872-2454873 & R & 33.04623 &  85 & 13.63736 & 0.003144 \\ \hline

    I1 & 2454824 & I & 63.01641 & 107 &  0.05715 & 0.002475 \\ \hline
    I2 & 2454825 & I & 37.51954 & 109 & -0.11376 & 0.003702 \\ \hline
    I3 & 2454826 & I & 48.21065 & 129 &  0.015481 & 0.001534 \\ \hline
    I4 & 2454828 & I & 29.17502 &  51 &  0.300137 & 0.002157 \\ \hline
    I5 & 2454830 & I & 44.49193 &  84 &  0.047845 & 0.000809 \\ \hline
    I6 & 2454866-2454867 & I & 63.7564 & 125 &  1.357992  & 0.001129 \\ \hline
    I7 & 2454871-2454872 & I & 32.23456 & 102 &  0.054029 & 0.005672 \\ \hline
    \caption{Observational data, with $B \to 440 nm$, $R \to 630 nm$, $V\to 550 nm$, $I\to 900 nm$.} \label{table:variabilityData}
\end{longtable}

\begin{longtable}[!t]{ | p{0.55cm} | p{0.4cm} | p{2.4cm} | p{2.4cm} | p{2.4cm} | p{2cm} | p{1cm}|}
    \hline
    &  & $\theta _1$ & $\theta _2$ & $\theta _3$ & $\theta _4$& $p_B$ \\ \hline
    \multirow{3}{*} {B1}         & $H_1$ & 0.782 [1.415]                 & 11.427 [21.173]  & -19.091 [11.405] & -34.85 [14.401] & 1 \\
                                 & $H_2$ & \emph{\textbf{1.821}} [0.155] & -11.765 [0.633]  & -                & -               & 1 \\
                                 & $H_3$ & 1.923 [0.662]                 & -19.949 [24.702] & 1.222 [11.639]   & -               & 0.999 \\ \hline

    \multirow{2}{*}{\textbf{B2}} & $H_0$ & 2.517 [0.333]                 & -12.214 [0.962]  & -22.583 [0.349]  & -               & 0.501 \\
                                 & $H_2$ & 1.828 [0.116]                 & -13.802 [0.498]  & -                & -               & 0.193 \\ \hline

    \multirow{2}{*}{B3}          & $H_0$ & \emph{\textbf{2.184}} [0.557] & -12.081 [1.641]  & -27.685 [7.306]  & -               & 0.96 \\
                                 & $H_2$ & 1.877 [0.262]                 & -12.852 [1.02]   & -                & -               & 0.683 \\ \hline

    \multirow{3}{*}{B4}          & $H_0$ & 2.15 [0.304]                  & -11.278 [0.902]  & -22.029 [1.615]  & -               & 1 \\
                     & $H_1$ & 2.137 [0.588]                 & -9.011 [6.632]   & -2.083 [10.563]  & -21.828 [1.34]  & 0.999 \\
                                 & $H_2$ & \emph{\textbf{1.842}} [0.124] & -12.058 [0.523]  & -                & -               & 0.945 \\ \hline

    \multirow{3}{*}{B5}          & $H_0$ & 2.286 [0.412]                 & -9.341 [1.122]   & -18.867 [0.871]  & -               & 0.938 \\
                     & $H_1$ & 3.637 [2.305]                 & -3.87 [17.143]   & -5.56 [7.76]     & -19.697 [1.701] & 0.912 \\
                                 & $H_2$ & \emph{\textbf{1.654}} [0.13]  & -10.862 [0.523]  & -                & -               & 0.997 \\ \hline

    \multirow{2}{*}{B6}      & $H_0$ & \emph{\textbf{2.639}} [0.58]  & -11.141 [1.273]  & -19.78 [0.334]   & -               & 0.914 \\
                                 & $H_2$ & 1.53 [0.143]                  & -13.228 [0.557]  & -                & -               & 0.397 \\ \hline
                              & & & & & &\\ \hline
    \multirow{2}{*}{V1}          & $H_0$ & 2.372 [0.389]                 & -8.82 [1.313]    & -22.231 [3.229]  & -               & 1 \\
                                 & $H_2$ & \textbf{\emph{1.964}} [0.143] & -10.061 [0.664]  & -                & -               & 0.994 \\ \hline

    \multirow{2}{*}{V2}          & $H_0$ & 2.271 [0.404]                 & -9.316 [1.328]   & -20.777 [1.397]  & -               & 1 \\
                                 & $H_2$ & \emph{\textbf{1.822}} [0.158] & -10.622 [0.714]  & -                & -               & 1 \\ \hline

    \multirow{4}{*}{V3}          & $H_0$ & 2.026 [0.243]                 & -12.837 [0.972]  & -44.782 [15.266] & -               & 0.926 \\
                 & $H_1$ & 3.195 [1.712]                 & 4.575 [5.77]     & -14.488 [2.021]  & -28.615 [4.354] & 0.947 \\
                                 & $H_2$ & \textbf{\emph{2.012}} [0.212] & -12.857 [0.886]  & -                & -               & 0.857 \\
                 & $H_3$ & 2.76 [2.059]                  & 3.792 [76.601]   & 0.24 [25.477]    & -               & 0.882 \\ \hline

    \multirow{3}{*}{V4}          & $H_0$ & 2.28 [0.388]                  & -10.145 [1.278]  & -25.1 [5.594]    & -               & 1 \\
                 & $H_1$ & 1.558 [0.896]                 & -0.124 [41.476]  & -9.322 [7.371]   & -22.437 [1.019] & 1 \\
                                 & $H_2$ & \emph{\textbf{1.965}} [0.158] & -11.067 [0.709]  & -                & -               & 1 \\ \hline

    \multirow{3}{*}{V5}          & $H_0$ & 1.932 [0.341]                 & -11.664 [1.044]  & -23.652 [5.714]  & -               & 1 \\
                 & $H_1$ & 3.583 [3.026]                 & -1.417 [4.747]   & -9.868 [4.274]   & -20.371 [0.567] & 1 \\
                                 & $H_2$ & \emph{\textbf{1.59}} [0.137]  & -12.579 [0.556]  & -                & -               & 0.992 \\ \hline

    \multirow{2}{*}{\textbf{V6}} & $H_0$ & 3.218 [0.401]                 & -10.434 [0.935]  & -22.473 [0.283]  & -               & 0.409 \\
                                 & $H_2$ & 2.237 [0.116]                 & -12.347 [-0.488] & -                & -               & 0.042 \\ \hline

    \multirow{2}{*}{\textbf{V7}} & $H_0$ & 2.76 [0.424]                  & -12.099 [1.024]  & -22.186 [0.257]  & -               & 0.48 \\
                                 & $H_2$ & 1.692 [0.106]                 & -14.297 [0.46]   & -                & -               & 0.014 \\ \hline

    \multirow{2}{*}{V8}          & $H_0$ & 2.914 [1.388]                 & -11.22 [2.894]   & -19.98 [0.946]   & -               & 0.998 \\
                                 & $H_2$ & \emph{\textbf{1.083}} [0.199] & -14.964 [0.789]  & -                & -               & 0.83 \\ \hline

    \multirow{2}{*}{V9}          & $H_0$ & \emph{\textbf{2.331}} [0.266] & -12.314 [0.842]  & -22.683 [0.307]  & -               & 0.923 \\
                                 & $H_2$ & 1.683 [0.103]                 & -14.01 [0.475]   & -                & -               & 0.537 \\ \hline

    \multirow{2}{*}{V10}         & $H_0$ & 1.636 [0.633]                 & -15.595 [1.454]  & -23.307 [2.94]   & -               & 0.776 \\
                                 & $H_2$ & \emph{\textbf{1.06}} [0.155]  & -16.763 [0.599]  & -                & -               & 0.718 \\ \hline

    \multirow{2}{*}{V11}         & $H_0$ & 1.788 [0.166]                 & -13.235 [0.618]  & -28.954 [6.43]   & -               & 0.536 \\
                                 & $H_2$ & \emph{\textbf{1.722}} [0.097] & -13.423 [0.447]  & -                & -               & 0.661 \\ \hline

    \multirow{3}{*}{V12}         & $H_0$ & \emph{\textbf{2.65}} [0.377] & -10.146 [1.101]  & -20.554 [0.27]   & -               & 0.923 \\
                 & $H_1$ & 3.167 [0.988]                 & -0.139 [2.438]   & -9.634 [3.319]   & -20.473 [0.225] & 0.984 \\
                                 & $H_2$ & 1.664 [0.114]                 & -12.653 [0.478]  & -                & -               & 0.662 \\ \hline

    \multirow{2}{*}{V13}         & $H_0$ & 1.816 [0.276]                 & -10.514 [0.829]  & -25.257 [5.452]  & -               & 0.952 \\
                                 & $H_2$ & \emph{\textbf{1.708}} [0.145] & -10.75 [0.582]   & -                & -               & 0.969 \\ \hline

    \multirow{2}{*}{\textbf{V14}}& $H_0$ & 2.504 [0.434]                 & -11.444 [1.095]  & -20.488 [0.507]  & -               & 0.359 \\
                                 & $H_2$ & 1.665 [0.141]                 & -13.144 [0.55]   & -                & -               & 0.041 \\ \hline
        & & & & & &\\ \hline
    \multirow{2}{*}{R1}          & $H_0$ & 2.299 [0.343]                 & -9.162 [1.192]   & -22.296 [3.029]  & -               & 1 \\
                                 & $H_2$ & \emph{\textbf{1.971}} [0.143] & -10.151 [0.653]  & -                & -               & 0.994 \\ \hline

    \multirow{2}{*}{R2}          & $H_0$ & 1.963 [0.169]                 & -10.265 [0.758]  & -33.563 [4.956]  & -               & 1 \\
                                 & $H_2$ & \emph{\textbf{1.957}} [0.164] & -10.288 [0.735]  & -                & -               & 1 \\ \hline

    \multirow{3}{*}{R3}          & $H_0$ & 2.548 [1.603]                 & -11.829 [4.261]  & -68.786 [52.262] & -               & 0.999 \\
                                 & $H_2$ & \emph{\textbf{1.906}} [0.228] & -13.515 [0.968]  & -                & -               & 1 \\
                                 & $H_3$ & 1.78 [2.459]                  & -1.949 [14.328]  & -13.67 [6.54]    & -               & 0.994 \\ \hline

    \multirow{2}{*}{R4}          & $H_0$ & 2.287 [0.428]                 & -10.186 [1.397]  & -22.434 [2.601]  & -               & 1 \\
                                 & $H_2$ & \emph{\textbf{1.854}} [0.158] & -11.444 [0.699]  & -                & -               & 0.999 \\ \hline

    \multirow{3}{*}{R5}          & $H_0$ & 1.715 [0.217]                 & -12.345 [0.717]  & -29.301 [9.384]  & -               & 0.934 \\
                 & $H_1$ & 1.988 [1.533]                 & -8.417 [18.501]  & -9.563 [3.639]   & -24.388 [3.974] & 0.93 \\
                                 & $H_2$ & \emph{\textbf{1.654}} [0.151] & -12.466 [0.621]  & -                & -               & 0.951 \\ \hline

    \multirow{2}{*}{R6}          & $H_0$ & \emph{\textbf{2.413}} [0.277] & -11.689 [0.76]   & -24.214 [2.621]  & -               & 0.876 \\
                                 & $H_2$ & 2.178 [0.121]                 & -12.165 [0.505]  & -                & -               & 0.256 \\ \hline

    \multirow{2}{*}{R7}          & $H_0$ & \emph{\textbf{2.389}} [0.318] & -12.693 [0.869]  & -22.236 [0.323]  & -               & 0.804 \\
                                 & $H_2$ & 1.671 [0.108]                 & -14.255 [0.465]  & -                & -               & 0.312 \\ \hline

    \multirow{2}{*}{R8}          & $H_0$ & 2.767 [1.933]                 & -13.789 [4.617]  & -24.024 [5.106]  & -               & 0.929 \\
                                 & $H_2$ & \emph{\textbf{1.51}} [0.241]  & -13.792 [0.943]  & -                & -               & 0.859  \\ \hline

    \multirow{2}{*}{R9}          & $H_0$ & 1.934 [0.274]                 & -13.187 [0.879]  & -25.392 [3.993]  & -               & 0.983 \\
                                 & $H_2$ & \emph{\textbf{1.666}} [0.107] & -13.941 [0.492]  & -                & -               & 0.923 \\ \hline

    \multirow{3}{*}{R10}         & $H_0$ & 2.668 [0.944]                 & -13.513 [1.909]  & -21.319 [0.297]  & -               & 0.606 \\
                 & $H_1$ & 1.696 [1.723]                 & 57.207 [119.484] & -84.994 [161.649]& -21.591 [1.058] & 0.604 \\
                                 & $H_2$ & \emph{\textbf{1.107}} [0.143] & -16.565 [0.56]   & -                & -               & 0.67 \\ \hline

    \multirow{2}{*}{\textbf{R11}}& $H_0$ & 1.724 [0.114]                 & -13.101 [0.512]  & -54.669 [17.637] & -               & 0.28 \\
                                 & $H_2$ & 1.717 [0.107]                 & -13.122 [0.49]   & -                & -               & 0.311 \\ \hline

    \multirow{3}{*}{R12}         & $H_0$ & \emph{\textbf{2.399}} [0.364] & -11.009 [1.073]  & -24.938 [6.496]  & -               & 0.962 \\
                 & $H_1$ & 2.567 [0.452]                 & -2.288 [2.48]    & -7.474 [3.559]   & -21.72 [0.511]  & 0.967 \\
                                 & $H_2$ & 2.006 [0.145]                 & -12.001 [0.585]  & -                & -               & 0.745 \\ \hline

    \multirow{2}{*}{R13}         & $H_0$ & 1.881 [0.245]                 & -10.344 [0.76]   & -25.802 [5.618]  & -               & 0.995 \\
                                 & $H_2$ & \emph{\textbf{1.78}}  [0.144] & -10.606 [0.578]  & -                & -               & 0.979 \\ \hline

    \multirow{2}{*}{R14}         & $H_0$ & \emph{\textbf{2.517}} [0.397] & -11.687 [1.049]  & -21.737 [0.678]  & -               & 0.959 \\
                                 & $H_2$ & 1.944 [0.146]                 & -12.9 [0.574]    & -                & -               & 0.774 \\ \hline
        & & & & & &\\ \hline
    \multirow{3}{*}{I1}          & $H_0$ & 1.364 [0.202]                 & -2.158 [0.727]   & -20.594 [7.576]  & -               & 0.991 \\
                 & $H_1$ & 2.07 [1.577]                  & 1.232 [4.893]    & -1.496 [2.264]   & -12.817 [3.63]  & 0.982 \\
                                 & $H_2$ & \emph{\textbf{1.333}} [0.156]        & -2.239 [0.64]    & -                & -               & 0.985 \\ \hline

    \multirow{2}{*}{\textbf{I2}} & $H_0$ & 2.361 [0.312]                 & -1.834 [0.893]   & -11.504 [0.428]  & -               & 0.282 \\
                                 & $H_2$ & 1.775 [0.119]                 & -3.099 [0.502]   & -                & -               & 0.258 \\ \hline

    \multirow{2}{*}{\textbf{I3}} & $H_0$ & 2.728 [0.416]                 & 1.421 [1.06]     & -7.975 [0.188]   & -               & 0.361 \\
                                 & $H_2$ & 1.395 [0.094]                 & -1.406 [0.415]   & -                & -               & 0.033 \\ \hline

    I4               & $H_2$ & \emph{\textbf{1.405}} [0.244] & -6.419 [0.942]   & -                & -               & 0.747 \\ \hline

    \multirow{3}{*}{\textbf{I5}} & $H_0$ & 1.943 [0.979]                 & -3.732 [1.843]   & -10.396 [1.627]  & -               & 0.175 \\
                 & $H_1$ & 3.144 [2.228]                 & -32.259 [374.247]& 31.418 [278.444] & -8.946 [0.711]  & 0.19 \\
                                 & $H_2$ & 0.903 [0.156]                 & -5.634 [0.604]   & -                & -               & 0.151 \\ \hline

    \multirow{2}{*}{I6}          & $H_2$ & 0.376 [0.167]                 & -13.277 [0.688]  & -                & -               & 0.622 \\
                 & $H_3$ & \emph{\textbf{1.702}} [2.878] & 6.767 [4.691]    & -10.319 [0.436]  & -               & 0.949 \\ \hline

    \multirow{2}{*}{I7}          & $H_0$ & 2.229 [0.323]                 & 0.52 [0.92]      & -8.81 [0.93]     & -               & 0.863 \\
                                 & $H_2$ & \emph{\textbf{1.722}} [0.133] & -0.625 [0.543]   & -                & -               & 0.952 \\ \hline
        \caption{Results of spectral analysis.} \label{table:spectral}
\end{longtable}

\end{onecolumn}


\begin{thebibliography}{}
  \bibitem{art:9a} Azarnia, G., Webb, J., Pollock, J.: 2005, I.A.P.P.P. Communications 101, 1
  \bibitem{art:11b} Balbus, S., Hawley, J.: 1991, ApJ 376, 214
  \bibitem{art:28b} Balbus, S., Hawley, J.: 1992a, ApJ 392, 662
  \bibitem{art:26b} Balbus, S., Hawley, J.: 1992b, ApJ 400, 610
  \bibitem{art:18b} Balbus, S., Hawley, J.: 1998, RevModpHys 70, 1
  \bibitem{art:8b} Begelman, M., Rees, J., Sikora, M.: 1994, ApJ 429, L57
  \bibitem{art:21b} Blaes, O.: 2002, Euro Summer School - NATO advanced study institute, 139
  \bibitem{art:26b} Balbus, S., Hawley, J.: 1992, ApJ 400, 610
  \bibitem{art:25c} Carini, M., Walter, R., Hopper, L.: 2011, ApJ 141, 49
  \bibitem{art:24c} Chandra, S., Baliyan, K., Ganesh, S., Joshi, U.: 2011,
  ApJ 731, 118
  \bibitem{} Fan, J.H., Tao, J., Qian, B.C., Liu, Y., Yang, J.H.,
  Pi, F.P., Xu, W.: 2011, RAA 11, 1311
  \bibitem{art:22h} Hawley, J., Balbus, S.: 1991, ApJ 376, 223
  \bibitem{art:29h} Hawley, J., Balbus, S.: 1992, ApJ 400, 595
  \bibitem{art:18h} Hawley, J., Gammie, C., Balbus, S.: 1995, ApJ 440, 472
  \bibitem{art:7k} Kirk, J., Mastichiadis, A.: 1992, Nature 360, 135
  \bibitem{art:34k} Kraus, A., Quirrenbach, A., Lobanov, A., et al: 1999, AA 344, 807
  \bibitem{art:1k} Krichbaum, T., Kraus, A., Fuhrman, L., Cimo, G., Witzel, A.: 2002, PASA 19, 14
  \bibitem{book:13} Krolik, J.: 1999, Active Galactiv Nuclei. From the central black hole to the galactiv environment, Princeton new Jersey: Princeton University Press
  \bibitem{art:14l} Lazarian, A., Vishniac, E.: 1999, ApJ 517, 700
  \bibitem{art:11l} Leung, C., Wei, J., Harko, T., Kovacs, Z.:
  2011a, J. Astrophys. Astr. 32, 189
  \bibitem{art:12l} Leung, C., Wei, J., Kong, A., Kovacs, Z., Harko, T.: 2011b,
  RAA 11, 1031L
  \bibitem{art:15m} Mineshige, S., Ouchi, N.B., Nishimori, H.: 1994, PASJ 46, 97
  \bibitem{art:16m} Mineshige, S., Takeuchi, N., Nishimori, H.: 1994, ApJ 435, L125
  \bibitem{art:30p} Papaloizou, J., Lin, D.: 1995, ANAA 33, 505
  \bibitem{art:3p} Poon, H., Fan, J., Fu, J.: 2009, ApJS 185, 511
  \bibitem{art:12q} Qian, B., Tao, J., Fan, J.: 2002, AJ 123, 678
  \bibitem{art:4q} Qian, S.: 1995, Chin.AA 19, 69
  \bibitem{art:7q} Qian, S., Krichbaum, T., Witzel, A., Zensus, J., Zhang, X.: 2006, Chin.AA 6, 530
  \bibitem{art:9q} Qian, S., Wegner, L., Witzel, A., Krichbaum, T.: 1996a, Chin.AA 20, 15
  \bibitem{art:10q} Qian, S.,Witzel, A., Kraus, A., Krichbaum, T., Britzen, S.: 1996b, ASp Conf. Series 100, 55
  \bibitem{art:3q} Quirrenbach, A., Witzel, A., Kirchbaujm, T.P., et al: 1992, AA 258, 279
  \bibitem{art:2q} Quirrenbach, A., Witzel, A., Wagner, S., et al: 1991, ApJ 372, L71
  \bibitem{art:15r} Raiteri, C., Villata, M., Tosti, G., et al: 2003, AA 402, 151
  \bibitem{art:12s} Shakura, N., Sunyaev, R.: 1973, AA 24, 337
  \bibitem{art:26s} Shakura, N., Sunyaev, R.: 1976, MNRAS 175, 623
  \bibitem{art:4v} Vaughan, S.: 2010, MNRAS 402, 307
  \bibitem{art:3v} Villata, M., Raiteri, C., Larionov, V., et al: 2008, AA 481, L79
  \bibitem{art:2w} Wagner, S., Sanchez-Pons, F., Quirrenbach, A., Witzal, A.: 1990, AA 235, L1
  \bibitem{art:5w} Wagner, S., Witzel, A.: 1995, ARAA 33, 163
  \bibitem{art:11w} Wagner, S., Witzel, A., Heidt, A., et al: 1996, AA 224, L9
  \bibitem{art:4w} Wambsganss, J., Schneider, P., Quirrenbach, A., Witzel, A.: 1989, AA 224, L9
  \bibitem{art:1y} Yonehara, A., Mineshige, S., Welsh, W.: 1997, ApJ 486, 388
\end{thebibliography}
\end{document}